\documentclass[conference]{IEEEtran}
\IEEEoverridecommandlockouts
\usepackage{cite}
\usepackage{amsmath,amssymb,amsfonts}
\usepackage{algorithmic}
\usepackage{graphicx}
\usepackage{textcomp}
\usepackage{xcolor}
\def\BibTeX{{\rm B\kern-.05em{\sc i\kern-.025em b}\kern-.08em
    T\kern-.1667em\lower.7ex\hbox{E}\kern-.125emX}}

\usepackage{graphicx}
\usepackage{booktabs}
\usepackage{multirow}
\usepackage{tabularx}
\usepackage{diagbox}
\usepackage{colortbl}
\usepackage{soul}
\usepackage{wrapfig}
\usepackage[caption=false]{subfig}
\usepackage[font=footnotesize,labelfont=bf]{caption}
\usepackage{dblfloatfix}
\usepackage{threeparttable} 
\usepackage{enumitem}

\usepackage{flushend}

\begin{document}

\title{StepGrade: Grading Programming Assignments with Context-Aware LLMs\\
}

\author{\IEEEauthorblockN{Mohammad Akyash, Kimia Zamiri Azar, Hadi Mardani Kamali}
\IEEEauthorblockA{\textit{Department of Electrical and Computer Engineering (ECE), University of Central Florida, Orlando, FL 32816, USA} \\
\{mohammad.akyash, azar, kamali\}@ucf.edu}
}

\maketitle

\begin{abstract}

Grading programming assignments is a labor-intensive and time-consuming process that demands careful evaluation across multiple dimensions of the code. To overcome these challenges, automated grading systems are leveraged to enhance efficiency and reduce the workload on educators. Traditional automated grading systems often focus solely on correctness, failing to provide interpretable evaluations or actionable feedback for students. This study introduces StepGrade, which explores the use of Chain-of-Thought (CoT) prompting with Large Language Models (LLMs) as an innovative solution to address these challenges. Unlike regular prompting, which offers limited and surface-level outputs, CoT prompting allows the model to reason step-by-step through the interconnected grading criteria, i.e., functionality, code quality, and algorithmic efficiency, ensuring a more comprehensive and transparent evaluation. This interconnectedness necessitates the use of CoT to systematically address each criterion while considering their mutual influence. To empirically validate the efficiency of StepGrade, we conducted a case study involving 30 Python programming assignments across three difficulty levels (easy, intermediate, and advanced). The approach is validated against expert human evaluations to assess its consistency, accuracy, and fairness. Results demonstrate that CoT prompting significantly outperforms regular prompting in both grading quality and interpretability. By reducing the time and effort required for manual grading, this research demonstrates the potential of GPT-4 with CoT prompting to revolutionize programming education through scalable and pedagogically effective automated grading systems.

\end{abstract}


\section{Introduction}

Grading programming assignments is a cornerstone of computer science education, offering students valuable feedback to enhance their coding proficiency and deepen their understanding of programming principles \cite{caiza2013programming}. This process not only reinforces core concepts but also helps identify areas where students may need additional support. However, as the number of students increases, the task of grading becomes more complex and time-intensive, demanding scalable solutions to maintain the quality of feedback \cite{messer2024automated}.

The grading challenge is further amplified in Massive Open Online Courses (MOOCs), where thousands of students participate in a single session \cite{bey2018comparison}. While MOOCs democratize access to high-quality education globally, their scalability introduces significant barriers to effective assessment \cite{miller2014functional}. Manual grading becomes impractical due to the sheer volume of submissions, and peer grading, while scalable, often results in inconsistent evaluations \cite{gamage2021peer}. Traditional automated grading systems have been developed to streamline this process, but they often focus narrowly on output correctness. This emphasis overlooks critical aspects such as code readability, effective error handling, and the delivery of insightful, pedagogically meaningful feedback \cite{albreiki2021systematic, autolab2023, 10.1145/1384271.1384371}. Early machine learning (ML)-based systems marked a significant advancement in automating the grading process, leveraging techniques such as feature extraction and classification to evaluate student code \cite{albreiki2021systematic, pallathadka2023classification, srikant2014system, geigle2016exploration}. These systems demonstrated the feasibility of automating specific aspects of grading, such as identifying patterns in coding errors, clustering common mistakes, and predicting student performance based on predefined criteria. However, their reliance on handcrafted features, and token-based representations, limited their adaptability to the diverse and dynamic nature of programming assignments.

LLMs have revolutionized diverse fields, from advancing chip design \cite{akyash2025simeval, wang2024llms} and strengthening hardware security \cite{akyash2024evolutionary, akyash2024selfhwdebug} to driving breakthroughs in drug discovery \cite{chakraborty2023artificial} and refining financial modeling \cite{zhao2024revolutionizing}. Their impact extends to education, where they are reshaping the way programming assignment are graded \cite{10.1145/3643795.3648375}. Their scalability makes them particularly well-suited for MOOCs, ensuring consistent and timely evaluations that enhance learning outcomes while significantly reducing instructors' workload \cite{gabbay2024combining}.

Among LLMs, GPT-4 \cite{achiam2023gpt} has demonstrated exceptional capabilities in understanding, analyzing, and generating code, making it highly suitable for automated grading tasks \cite{liu2024your}. Regular prompting \cite{brown2020language}, a commonly used approach with LLMs, involves providing concise queries to the model to elicit direct responses. While this method can be effective for basic grading tasks or straightforward assessments, it often falls short when applied to the complex, multi-faceted nature of programming assignment evaluation.

Regular prompting has two major limitations when applied to grading programming assignments: 

\noindent \ul{\textit{(i) Lack of contextual reasoning}} --- Regular prompting treats grading criteria such as \textit{functionality}, \textit{code quality}, and \textit{algorithmic efficiency} as isolated tasks rather than interconnected components. This fragmented evaluation prevents the model from recognizing how these aspects influence each other, leading to feedback that lacks depth and coherence. 

\noindent \ul{\textit{(ii) Inconsistent and opaque evaluations}} --- Without systematic reasoning, regular prompting often produces generic or inconsistent assessments. It may overlook critical aspects such as edge cases, inefficiencies, or best practices. Additionally, the evaluation process lacks transparency, as it does not provide clear reasoning behind the assigned grade, making it difficult for educators/students to trust and learn from the feedback.

\begin{figure}[t]
    \centering
    \includegraphics[width=0.9\columnwidth]{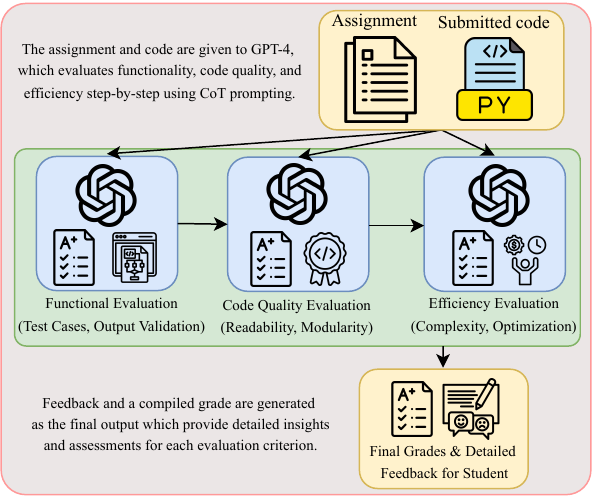}
    \caption{The Overall Grading Flow in StepGrade Framework: Sequential Evaluation of Key Grading Criteria, i.e., Functionality, Code Quality, and Efficiency using CoT Prompting, with Detailed Feedback and Grades Outputs.}
    \vspace{-15pt}
    \label{fig:stepgrade_framework}
\end{figure}

To address the limitations of regular prompting in grading programming assignments, we propose \textbf{StepGrade}, a framework that utilizes Chain-of-Thought (CoT) prompting \cite{wei2022chain} to guide the model through a structured, step-by-step grading process. The StepGrade framework is shown in Fig. \ref{fig:stepgrade_framework}, which follows a structured evaluation for the use of COT-based LLM for grading. Our key highlights are as follows:

\begin{itemize}[leftmargin=*]
    \item \textbf{StepGrade Framework:} We introduce \textbf{StepGrade}, a novel grading framework that leverages CoT prompting to systematically evaluate key grading criteria, i.e., \textit{functionality}, \textit{code quality}, and \textit{algorithmic efficiency}. Unlike regular prompting, which often leads to fragmented or inconsistent evaluations, StepGrade ensures that these criteria are analyzed in an interconnected and coherent manner.
    
    \item \textbf{Enhanced Transparency and Interpretability:} StepGrade improves grading transparency by explicitly revealing the model's reasoning process in a verbose way. By breaking down the evaluation into intermediate steps, our framework provides more interpretable and actionable feedback, providing deeper insights for both students and educators.
    
    \item \textbf{Empirical Evaluation:} As a case study, we conduct a comprehensive evaluation of StepGrade using 30 Python programming assignments across three difficulty levels: \textit{Easy}, \textit{Intermediate}, and \textit{Advanced}. Each assignment is systematically assessed based on defined criteria, with human evaluations serving as the reference standard.
    
    \item \textbf{Improved Grading Accuracy:} Our results show that StepGrade consistently outperforms regular prompting in grading accuracy across all difficulty levels. The grades assigned by StepGrade align more with human evaluations, which demonstrates its effectiveness in automated grading.
\end{itemize}

\section{Related Works}

The evaluation of programming assignments has been extensively studied, with various approaches developed to address the challenges of grading at scale while maintaining quality. These methods can be broadly categorized into traditional systems, ML-based approaches, and recent advancements using LLMs. Each of these methods contributes unique insights but also reveals limitations that motivate further innovation.

\subsection{Traditional Methods for Programming Grading}

Early grading systems, such as Autolab \cite{autolab2023} and Web-CAT \cite{10.1145/1384271.1384371}, laid the foundation for scalable programming assignment evaluation by employing predefined test cases to assess functional correctness. These systems demonstrated the potential for automation in grading but inherently focused on surface-level correctness, often neglecting essential programming attributes like code readability, algorithmic efficiency, and robustness. Similarly, tools like Javabrat \cite{patil2010automatic}, designed for Java and Scala within platforms such as Moodle, follow this correctness-centered paradigm, which limits their ability to address broader pedagogical goals.

To overcome some of these limitations, semantic analysis techniques like Latent Semantic Analysis (LSA) \cite{zen2011using} were explored. By comparing student submissions to model solutions in a semantic vector space, LSA provided a more complex grading mechanism, ensuring consistent evaluation of structural similarities. However, these systems inherited challenges from traditional methods, such as difficulty in assessing execution order and logical flow. 

\subsection{ML-based Methods for Programming Grading}

Building on the limitations of traditional systems, machine learning (ML) methods introduced models capable of extending beyond predefined test cases. These methods excel at identifying error patterns, predicting student performance, and offering personalized feedback. For example, Control Flow Graphs (CFGs) and Data Dependency Graphs (DDGs) can be utilized \cite{srikant2014system} to evaluate program logic and programming practices against a rubric. Similarly, supervised and active learning techniques can be leveraged with feature extraction—such as token, similarity, and selection features—to predict grades while minimizing manual intervention \cite{geigle2016exploration}.

Despite their advantages, these methods rely heavily on handcrafted features and curated training datasets, which limit their adaptability to diverse/uncommon programming styles. Moreover, the reliance on expert-annotated data introduces scalability issues, particularly in large educational settings where rapid grading is essential. While ML methods improve over traditional ones by capturing a wider range of errors and inefficiencies, they fall short of providing comprehensive assessments that align with human evaluators' judgments.

\begin{figure*}[!t]
    \centering
    \includegraphics[width=0.95\textwidth]{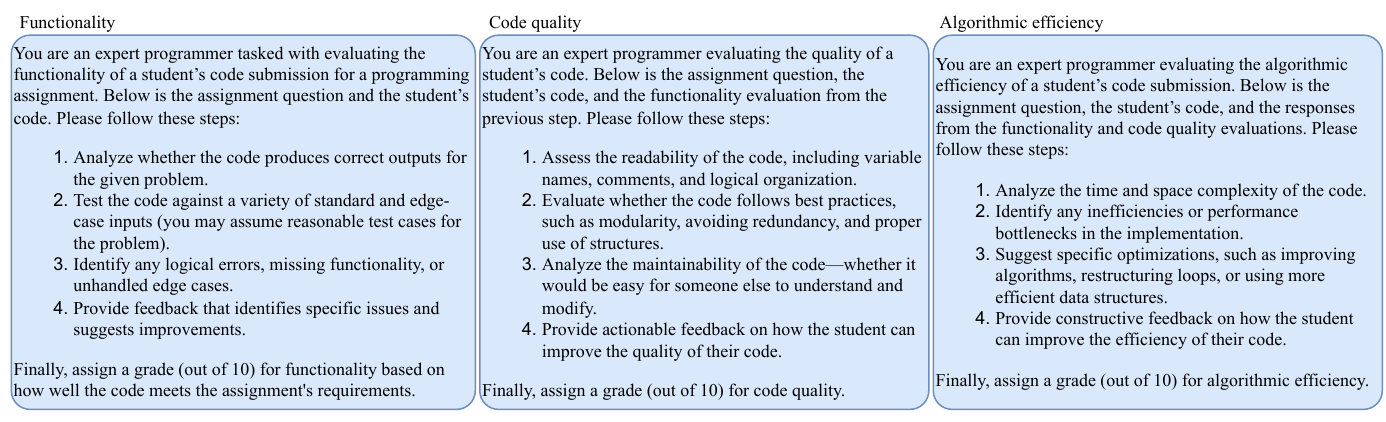}
    \caption{The Detailed Prompts utilized at Each Step of the COT prompting Process for Evaluating Functionality, Code Quality, and Algorithmic Efficiency.}
    \vspace{-15pt}
    \label{fig:prompts_cot}
\end{figure*}

\subsection{LLM-Based Methods for Programming Grading}

Recent advancements in LLMs, such as CodeBERT \cite{feng2020codebert}, Codex \cite{chen2021evaluating}, and ChatGPT-4 \cite{achiam2023gpt}, have revolutionized programming assignment evaluation by integrating syntax and semantic understanding into the grading process. These models show huge capabilities in identifying errors, suggesting improvements, and evaluating code functionality. However, their general-purpose design often limits their effectiveness in educational contexts, where feedback must align with instructional goals and consider a variety of grading criteria.

Several models have emerged to address this gap. GreAIter \cite{10.1145/3643795.3648375} is a ChatGPT-4-based rubric assessment model designed to enhance grading efficiency while using human oversight to manage false positives. Similarly, Dong and Liang \cite{10.1145/3674399.3674426} utilized CodeBERT and MiniLM-L6 to evaluate semantic alignment between code and assignment descriptions, achieving high accuracy for well-defined tasks. BeGrading \cite{yousef2024begrading} is also a fine-tuned LLM using a combination of real and synthetic datasets, yielding strong alignment with human evaluation.

While LLMs outperform traditional and ML-based systems, they often focus on isolated aspects like functionality or syntax, lacking holistic evaluation and adaptability to diverse assignments. To overcome these limitations, we propose CoT prompting, which emulates human reasoning by evaluating interconnected criteria—functionality, code quality, and efficiency—in context. CoT provides consistent, detailed, and objective feedback, ensuring scalability and alignment with instructional goals in diverse educational settings.

\section{Methodology: StepGrade}

Our method decomposes the task into smaller subtasks which enables consistent and human-like assessments. The following sections outline the principles of CoT prompting and its integration into our grading approach.

\subsection{Chain-of-Thought (CoT) Prompting}

CoT prompting enhances the reasoning capabilities of LLMs by breaking complex tasks into smaller, sequential steps, mirroring human problem-solving. Unlike regular prompting, which often produces shallow and fragmented responses, CoT emphasizes intermediate reasoning, enabling models to handle interconnected concepts effectively and provide thorough, context-aware solutions.

\subsubsection{How CoT Prompting Works}
CoT guides the model through each aspect of a problem, maintaining context and addressing dependencies. For instance, solving a mathematical problem involves identifying variables, forming relationships, and proceeding step-by-step to the solution. Similarly, CoT handles logical reasoning tasks by evaluating premises sequentially to ensure coherent and consistent outcomes.

\subsubsection{Comparison with Regular Prompting}

Regular prompting struggles with complex problems, often treating components in isolation and resulting in superficial outputs. In contrast, CoT excels by decomposing tasks into logical steps, enabling consistency, coherence, and interpretability. This structured approach makes CoT especially effective for multi-step problem-solving, reasoning, and evaluations requiring integration of multiple criteria.

\subsection{COT Implementation in StepGrade}

StepGrade evaluates programming assignments with COT-based LLMs in a step-by-step manner. It assesses functionality, quality, and efficiency, ensuring comprehensive and context-aware feedback. The process starts by providing the programming assignment to the LLM along with prompts to evaluate each aspect of the code. At each stage, the model:

\begin{enumerate}[leftmargin=*]

\item[(a)] Receives the assignment question, the submitted code, and contextual information from previous stages.

\item[(b)] Assigns a grade (from 1 to 10) for each evaluation criterion.

\item[(c)] Generates detailed feedback aimed at guiding the student to improve their submission (from instructor-perspective). 

\end{enumerate}

Each step of StepGrade is described below, outlining its inputs, outputs, and an illustrative example. Figure \ref{fig:prompts_cot} shows the prompts used at each stage.

\subsubsection{Step 1 --- Functional Evaluation}

The first stage assesses the functional correctness of the code. The LLM receives the assignment question, which provides the necessary context, and the student’s code. Using this information, the LLM evaluates whether the code produces the expected outputs for a variety of inputs, including edge cases, and identifies any logical flaws, incomplete functionality, or errors.

The output of this stage consists of a grade (on a scale of 10) representing the functional correctness and detailed feedback. The feedback highlights areas where the implementation succeeds and fails and offers suggestions to address specific shortcomings, such as handling missing edge cases or correcting logical errors.
For example, if the assignment involves implementing a sorting algorithm, the LLM might test the code with scenarios like empty arrays, already sorted arrays, and large datasets. The evaluation would confirm whether the outputs align with expectations and whether the code handles unusual cases effectively.

\subsubsection{Step 2 --- Code Quality Evaluation}

In the second stage, the LLM evaluates code quality, focusing on the readability, maintainability, and adherence to best practices in the submitted code. At this point, the LLM receives the assignment question, the student’s code and the functionality evaluation from the previous step. These inputs provide a contextual foundation which allows the LLM to assess how the implementation aligns with good coding standards.

At this stage, the LLM examines the clarity of variables' names, the logical structure of the code, and the presence of meaningful comments/documentation. It also evaluates whether the code is modular and follows best practices, such as avoiding redundancy and improving readability. The results of this evaluation include a grade for code quality and specific feedback. For instance, if the code lacks comments or enough clarity, the feedback would recommend replacing unclear names with descriptive ones and adding comments to clarify complex structures. This step ensures the code is not only correct but also easy to understand and maintain.

\subsubsection{Step 3 --- Algorithmic Efficiency Evaluation}

The final stage focuses on algorithmic efficiency, where the LLM assesses the computational performance of the submitted code. Here, the model receives the the assignment question, code submission along with the responses from both the functionality and code quality evaluations. By leveraging insights from the earlier steps, the LLM evaluates how efficiently the code solves the given problem. The evaluation examines time and space complexity, identifies performance bottlenecks, and suggests optimizations. For example, the LLM might detect that a nested loop could be replaced with a more efficient approach, such as using a hash table or adopting a divide-and-conquer algorithm. The output of this stage includes a grade for efficiency and feedback on improving performance. If the code employs a brute-force approach, the LLM might recommend switching to a more optimized algorithm, such as dynamic programming, to enhance efficiency and scalability.

\begin{table}[t]
\fontsize{6pt}{7pt}\selectfont
\centering
\caption{Description of the utilized Python Programming Assignments.}
\label{tab:assignment_description}
\setlength\tabcolsep{2pt} 
\begin{tabular}{@{} p{0.06\textwidth} p{0.12\textwidth} p{0.28\textwidth} @{} }
\toprule  
Difficulty & Assignment Title & Description \\
\cmidrule(r){1-1}\cmidrule(r){2-2}\cmidrule(r){3-3}
\multirow{10}{*}{Easy} 
& Calculate Factorial & Write a function to compute the factorial of a given number. \\
& Palindrome Checker & Develop a program to check if a given string is a palindrome. \\
& Fibonacci Sequence & Generate the Fibonacci sequence up to a given number. \\
& Prime Number Detector & Write a program to determine if a number is prime. \\
& Temperature Converter & Convert temperatures between °C, °F, and K. \\
& Basic Calculator & Calc. that performs ADD, SUB, MULT, and DIV. \\
& Reverse a String & Implement a function to reverse a given string. \\
& Sum of List Elements & Write a program to calculate the sum of all numbers in a list. \\
& Find Max and Min & Function to find the largest and smallest numbers in a list. \\
& Simple Interest Calc. & Calculate interest based on principal, rate, and time. \\
\cmidrule(r){1-1}\cmidrule(r){2-2}\cmidrule(r){3-3}
\multirow{10}{*}{Intermediate} 
& Matrix Multiplication & Perform matrix multiplication on two input matrices. \\
& Sorting Algorithms & Implement bubble sort and quicksort algorithms. \\
& Binary Search & Create a function to perform binary search on a sorted list. \\
& Word Frequency Count & Count the frequency of words in a given text file. \\
& Caesar Cipher & Encrypt and decrypt messages using a Caesar cipher. \\
& Armstrong No. Checker & A program to check if a number is an Armstrong number. \\
& File Handling & Read/write data from/to a text file and do file operations. \\
& Tower of Hanoi Solver & A recursive solution to the Tower of Hanoi problem. \\
& LCM and GCD Finder & Functions to compute the LCM and GCD of two numbers. \\
& Quad Eq. Solver & Solve quadratic equations using the discriminant formula. \\
\cmidrule(r){1-1}\cmidrule(r){2-2}\cmidrule(r){3-3}
\multirow{10}{*}{Advanced} 
& Graph Traversal & Traverse a graph using BFS and DFS. \\
& Dijkstra’s Algorithm & Dijkstra’s for the shortest path in a weighted graph. \\
& Knapsack Problem & Solve the 0/1 knapsack using dynamic programming. \\
& Sudoku Solver & Solve a Sudoku puzzle using backtracking. \\
& Web Scraper & Develop a web scraper to extract data from a website. \\
& Chatbot & Implement a simple chatbot using keyword-based responses. \\
& Stock Price Analysis & Analyze historical stock price for moving averages/trends. \\
& Database CRUD Opr. & Create, Read, Update, and Delete operations using SQLite. \\
& API Integration & Write a script to fetch and process data from a RESTful API. \\
& ML Model & A basic linear regression to predict outcomes based on input. \\
\bottomrule
\end{tabular}
\vspace{-15pt}
\end{table}

\begin{table*}[t]
\fontsize{6pt}{7pt}\selectfont
\centering
\caption{Lower Scoring Deviation from Human to Regular Prompting vs. CoT Prompting used in StepGrade.}
\label{tab:merged_results}
\setlength\tabcolsep{2pt} 
\begin{tabular}{@{} p{0.1\textwidth} p{0.2\textwidth} p{0.052\textwidth} p{0.052\textwidth} p{0.1\textwidth} p{0.052\textwidth} p{0.052\textwidth} p{0.1\textwidth} p{0.052\textwidth} p{0.052\textwidth} p{0.1\textwidth} @{} }
\toprule  
\multirow{2}{*}{Level} & \multirow{2}{*}{Assignment Title} & \multicolumn{3}{c}{Functionality} & \multicolumn{3}{c}{Code Quality} & \multicolumn{3}{c}{Algorithmic Efficiency} \\
\cmidrule(lr){3-5} \cmidrule(lr){6-8} \cmidrule(lr){9-11}
& & Human & Regular & \textbf{CoT (This work)} & Human & Regular & \textbf{CoT (This work)} & Human & Regular & \textbf{CoT (This work)} \\
\midrule
\multirow{11}{*}{Easy} 
 & Calculate Factorial & 9.5 & -0.5 & -0.5 & 8.5 & -0.8 & -0.4 & 9.0 & -0.8 & +0.1 \\
 & Palindrome Checker & 9.0 & -0.3 & -0.5 & 8.5 & +0.4 & -0.3 & 10.0 & -1.3 & -0.9 \\
 & Fibonacci Sequence & 7.5 & +0.8 & +0.4 & 8.0 & +1.0 & +0.5 & 7.5 & +1.4 & +1.2 \\
 & Prime Number Detector & 8.5 & -0.4 & +0.7 & 8.5 & -0.3 & +0.5 & 8.0 & 0.0 & +0.5 \\
 & Temperature Converter & 8.0 & +0.6 & +0.5 & 9.0 & -0.7 & -0.7 & 9.0 & -0.3 & +0.1 \\
 & Basic Calculator & 8.5 & -0.4 & +0.3 & 8.0 & +0.3 & +0.5 & 9.0 & -0.2 & 0.0 \\
 & Reverse a String & 10.0 & -0.8 & -0.9 & 9.0 & -0.6 & -0.2 & 8.5 & +0.4 & 0.0 \\
 & Sum of List Elements & 8.0 & -0.9 & -0.1 & 8.0 & +0.2 & +0.2 & 9.0 & -0.8 & +0.2 \\
 & Find Maximum and Minimum & 9.0 & -1.0 & 0.0 & 8.5 & 0.0 & -0.2 & 9.0 & -0.8 & +0.6 \\
 & Simple Interest Calculator & 9.0 & -0.9 & -0.5 & 9.5 & -1.0 & -0.5 & 8.5 & +0.1 & +0.5 \\
\cmidrule(r){2-11}
 & \textbf{Mean Absolute Error} & N/A & 0.66 & \textbf{0.44} & N/A & 0.53 & \textbf{0.40} & N/A & 0.61 & \textbf{0.41} \\
\cmidrule(r){1-11}
\multirow{11}{*}{Intermediate} 
 & Matrix Multiplication & 10.0 & -1.3 & -1.5 & 8.5 & -1.0 & +0.2 & 8.0 & -0.2 & +0.2 \\
 & Sorting Algorithms & 8.5 & -0.5 & -0.3 & 8.0 & -0.4 & +0.2 & 9.0 & -1.2 & -0.5 \\
 & Binary Search & 7.5 & +1.0 & +0.5 & 7.5 & +0.5 & +0.5 & 8.0 & -0.4 & 0.0 \\
 & Word Frequency Counter & 9.0 & -0.9 & -0.4 & 8.5 & -0.3 & 0.0 & 7.5 & +1.0 & +0.5 \\
 & Caesar Cipher & 8.5 & -0.2 & +0.1 & 8.0 & -0.4 & +0.5 & 8.5 & +0.5 & +0.1 \\
 & Armstrong Number Checker & 8.5 & -0.4 & -0.4 & 8.0 & +0.3 & 0.0 & 7.5 & +0.5 & +0.5 \\
 & File Handling & 7.5 & +0.4 & +0.5 & 7.5 & +0.2 & +0.5 & 7.5 & -0.3 & +0.4 \\
 & Tower of Hanoi Solver & 8.0 & +0.5 & +0.2 & 7.5 & +0.3 & +0.4 & 7.5 & -0.5 & 0.0 \\
 & LCM and GCD Finder & 8.5 & +0.5 & +0.3 & 8.0 & +0.5 & 0.0 & 9.0 & -0.1 & +0.2 \\
 & Quadratic Equation Solver & 7.0 & +1.4 & +1.0 & 8.0 & +0.5 & +0.1 & 8.5 & +0.4 & -0.2 \\
\cmidrule(r){2-11}
 & \textbf{Mean Absolute Error} & N/A & 0.71 & \textbf{0.52} & N/A & 0.44 & \textbf{0.24} & N/A & 0.51 & \textbf{0.26} \\
\cmidrule(r){1-11}
\multirow{11}{*}{Advanced} 
 & Graph Traversal & 7.5 & +1.5 & +1.2 & 7.5 & +1.1 & +0.8 & 8.5 & +0.5 & 0.0 \\
 & Dijkstra’s Algorithm & 7.5 & +1.0 & +1.1 & 7.5 & +1.2 & +0.5 & 7.5 & +0.7 & +0.5 \\
 & Knapsack Problem & 8.0 & 0.0 & +0.1 & 7.5 & +1.5 & +0.7 & 7.0 & +1.2 & +0.9 \\
 & Sudoku Solver & 9.0 & -0.5 & +0.3 & 8.5 & -0.2 & +0.1 & 9.0 & -0.5 & -0.7 \\
 & Web Scraper & 8.0 & +0.5 & +0.4 & 7.5 & +0.2 & +0.5 & 8.0 & +0.8 & +0.5 \\
 & Chatbot & 8.0 & -0.1 & +0.3 & 7.5 & +1.1 & +0.5 & 8.0 & +1.0 & +0.5 \\
 & Stock Price Analysis & 9.5 & -0.5 & -0.6 & 8.5 & -0.5 & +0.3 & 9.0 & -1.0 & -0.8 \\
 & Database CRUD Operations & 8.0 & +1.0 & +0.7 & 7.0 & +1.5 & +1.2 & 7.5 & +0.5 & +0.7 \\
 & API Integration & 8.5 & -0.5 & +0.4 & 6.5 & +1.5 & +1.2 & 7.5 & +0.5 & +0.5 \\
 & Machine Learning Model & 7.5 & +0.6 & +0.5 & 8.5 & -0.7 & +0.4 & 7.5 & +0.5 & + 0.7 \\
\cmidrule(r){2-11}
 & \textbf{Mean Absolute Error} & N/A & 0.62 & \textbf{0.56} & N/A & 0.95 & \textbf{0.62} & N/A & \textbf{0.54} & 0.58 \\
\bottomrule
\end{tabular}
\vspace{-15pt}
\end{table*}

\subsection{Why CoT Prompting is Effective}

Functionality, code quality, and algorithmic efficiency are deeply interconnected aspects of programming assignments, and their evaluation requires a context-aware approach to ensure meaningful feedback. Functional correctness ensures the code achieves its intended purpose, but without quality and efficiency, even correct code can be difficult to maintain or inefficient in practice. Code quality impacts readability, modularity, and scalability, which in turn affect functionality and performance. Similarly, algorithmic efficiency underpins the practicality of the solution. CoT prompting leverages this interconnectedness by evaluating each aspect within its broader context, so it provides detailed, context-aware feedback, which supports deeper learning, helping students recognize trade-offs, optimize solutions, and write more robust programs.

\section{Experiments and Results}

This section provides an overview of the dataset used and the experimental setup employed to evaluate the performance of StepGrade. The experiments focused on comparing the alignment of LLM-predicted grades with human evaluations across functionality, code quality, and algorithmic efficiency.

\subsection{Dataset}

To evaluate the effectiveness of the StepGrade framework, we utilized a dataset of 30 Python programming assignments. These assignments were collected from real student submissions, providing a diverse and authentic set of tasks for testing. The dataset was designed to cover a broad spectrum of programming challenges and was categorized into three difficulty levels—Easy, Intermediate, and Advanced—based on the complexity of the problem, the algorithmic requirements, and the expected coding effort.

\textbf{Easy Assignments.}
These assignments involve fundamental programming concepts, such as loops, conditionals, basic mathematical computations, and simple data manipulations. Examples include calculating a factorial, reversing a string, and checking if a number is prime.

\textbf{Intermediate Assignments.}
Assignments at this level require a combination of algorithms, data structures, and moderate problem-solving skills. They include tasks such as implementing sorting algorithms, performing matrix multiplication, and solving the Tower of Hanoi problem.

\textbf{Advanced Assignments.}
Advanced-level tasks involve complex algorithms, optimization problems, or the use of advanced programming techniques. Examples include solving the 0/1 knapsack problem, implementing graph traversal algorithms, and building a basic linear regression model.

The titles and the short descriptions of all 30 assignments are summarized in Table \ref{tab:assignment_description} and it demonstrate the variety of challenges included in the dataset.

\subsection{Comparison of Regular Prompting and CoT Prompting}

\begin{figure}[t]
    \centering
    \includegraphics[width=0.48\textwidth]{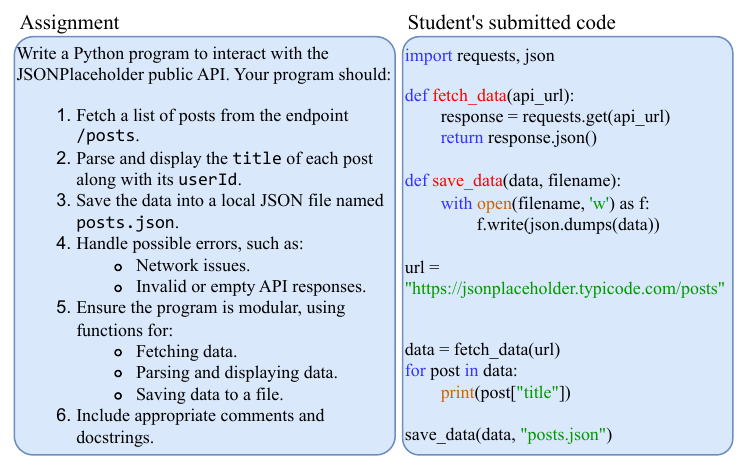}
    \caption{An Example of Assignments and the Code Submitted by the Student.}
    \vspace{-15pt}
    \label{fig:assignment_solution}
\end{figure}

\begin{figure}[!t]
    \centering
    \includegraphics[width=.48\textwidth]{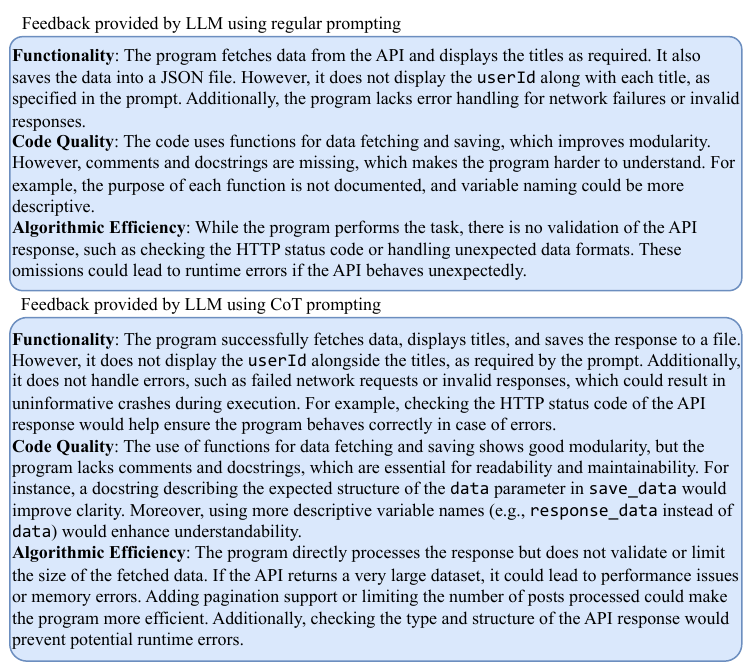}
    \caption{The Example Feedback provided by the LLM using Regular Prompting and CoT Prompting (StepGrade) Approaches.}
    \vspace{-10pt}
    \label{fig:feedback}
\end{figure}

\subsubsection{\textbf{Comparison of Grading}}  

We compared CoT prompting to regular prompting using human evaluations as the reference standard. Differences between LLM-predicted and human-assigned grades were analyzed, where positive values indicated more generous grading by the LLM and negative values indicated more conservative evaluations.

For functionality, CoT prompting consistently aligned better with human evaluations across all difficulty levels. As shown in Table \ref{tab:merged_results}, CoT had a smaller average deviation (0.44 vs. 0.66 for easy tasks, 0.52 vs. 0.71 for intermediate, and 0.56 vs. 0.62 for advanced). For code quality, CoT prompting also outperformed regular prompting, achieving lower deviations at all levels (0.40 vs. 0.53 for easy, 0.24 vs. 0.44 for intermediate, and 0.62 vs. 0.95 for advanced). For algorithmic efficiency, CoT prompting maintained its advantage, with deviations of 0.41 vs. 0.61 (easy), 0.26 vs. 0.51 (intermediate), and 0.58 vs. 0.54 (advanced). Overall, CoT prompting consistently outperformed regular prompting, delivering more accurate, context-aware assessments that closely aligned with human evaluations while providing actionable feedback.

\subsubsection{\textbf{Comparison of Feedback Generation}}
To evaluate StepGrade, we compared feedback from regular and CoT prompting on a student-submitted Python program (Fig. \ref{fig:assignment_solution}) that interacts with the JSONPlaceholder API. The assignment required fetching and displaying post titles with user IDs, saving data locally, and implementing error handling while maintaining modularity. Fig. \ref{fig:feedback} shows the generated feedback. For functionality, regular prompting noted missing user IDs and error handling but lacked specific corrective actions. CoT prompting suggested concrete improvements, such as checking HTTP status codes and validating response structures. For code quality, both methods identified the lack of comments and docstrings. Regular prompting offered a general critique, while CoT prompting provided actionable suggestions, such as adding docstrings, clarifying variable names (e.g., replacing \texttt{data} with \texttt{response\_data}), and improving readability. For algorithmic efficiency, regular prompting briefly mentioned missing validation but overlooked broader implications. CoT prompting highlighted potential inefficiencies, recommending pagination and response limits. It also recognized interdependencies, noting how missing error handling affected both functionality and performance.

\section{Conclusion}

This paper introduced StepGrade, a novel framework for grading programming assignments using LLMs. By leveraging CoT prompting, StepGrade systematically evaluates functionality, code quality, and algorithmic efficiency through interconnected, stepwise assessments. 
Through comprehensive experiments, we demonstrated the effectiveness of StepGrade across assignments of varying difficulty levels. CoT prompting consistently outperformed regular prompting, especially in handling complex tasks and maintaining alignment with human evaluations.

\bibliographystyle{IEEEtran}
\bibliography{refs}

\end{document}